# Unimpeded permeation of water through helium-leak-tight graphene-based membranes

R. R. Nair[1,2], H. A. Wu[1,3], P. N. Jayaram[2], I. V. Grigorieva[1], A. K. Geim[1,2]
[1]School of Physics and Astronomy, University of Manchester, Manchester M13 9PL, UK
[2]Centre for Mesoscience & Nanotechnology, University of Manchester, Manchester M13 9PL, UK
[3]CAS Key Laboratory, University of Science and Technology of China, Hefei, Anhui 230027, China

*Permeation through nanometer-pore materials has been attracting unwavering interest due to fundamental differences in governing mechanisms at macroscopic and molecular scales, the importance for filtration and separation techniques, and the crucial role played by selective molecular transport through cellular membranes. We have found that submicron-thick membranes made from graphene oxide are completely impermeable to liquids, vapors and gases, but allow unimpeded permeation of water ($H_2O$ permeates through the membranes at least $10^{10}$ times faster than He). We attribute these seemingly incompatible observations to a nearly frictionless flow of a monolayer of water through two dimensional capillaries formed by closely spaced graphene sheets. Diffusion of other molecules is blocked by the water that clogs the capillaries and by their reversible narrowing in low humidity.*

Despite being only one atom thick, graphene is believed to be impermeable to all gases and liquids [1]. This makes it tempting to exploit this material as a barrier film. Because of the ways graphene can currently be mass produced [2], films made from graphene oxide (GO) present a particularly interesting candidate. By using this graphene derivative, it is possible to make laminates, which are a collection of micron-sized graphene crystals forming an interlocked layered structure [3-5]. It resembles that of nacre and exhibits a great mechanical strength and flexibility even for films of submicron thickness [2-5]. In this report we investigate molecular permeation through such films.

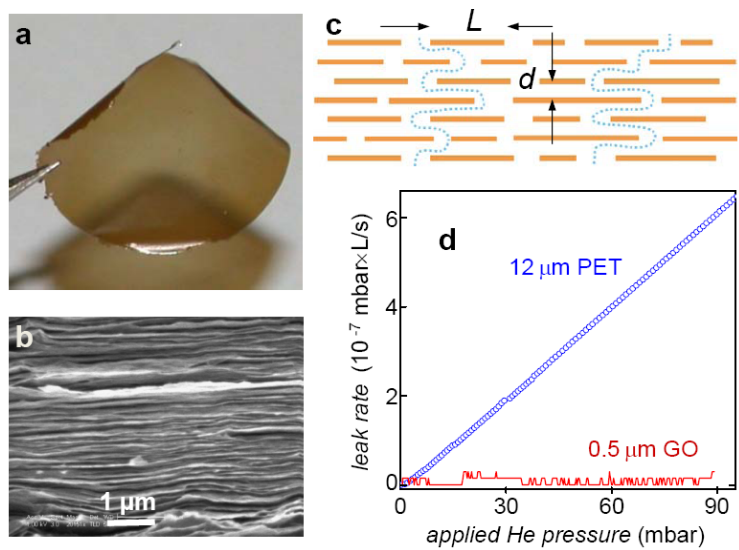

Figure 1. He-leak-tight GO membranes. **a** - Photo of a 1-μm-thick GO film peeled off a Cu foil. **b** - Electron micrograph of the film's cross-section. **c** - Schematic view for possible permeation through the laminates. Typical *L/d* is ~1000. **d** - Examples of He leak measurements for a freestanding sub-μm-thick GO membrane and a reference PET film (normalized per cm²).

Figure 1a shows an example of the studied GO membranes that, briefly, were prepared as follows [6]. First, Hummer's method was employed to obtain graphite oxide that was dispersed in water by sonication to make a stable suspension of GO crystals [3-5]. This suspension was used to produce laminates by spray- or spin-coating [5]. Scanning electron microscopy and X-ray analysis reveal that such GO films have a pronounced layered structure (Fig. 1b) and consist of crystals with typical sizes $L$ of a few µm, which are separated by a typical distance $d \approx 1$nm [3-5]. For permeation experiments, Cu foils of several cm in diameter were uniformly covered with the GO laminates. Then, we chemically etched Cu to produce apertures of diameter $D \approx 1$cm fully covered by freestanding GO films. Finally, a metal container was sealed by using the Cu disks. We studied membranes with thicknesses $h$ from 0.1 to 10 µm. Even sub-µm-thick membranes were strong enough to withstand a differential pressure $\Delta P$ up to 100 mbar.

As an initial test, we filled the containers with various gases under a small overpressure (<100 mbar) and recorded its changes over a period of several days. No noticeable reduction in $\Delta P$ was observed for any tested gas including He, hydrogen, nitrogen and Ar. This allowed an estimate for the upper limit on their permeation rates $Pr$ as $\approx 10^{-11}$ g/cm$^2 \times$s$\times$bar, which is close to the value reported for micron-sized "balloons" made from continuous graphene monolayers [1]. In an alternative approach, we used mass spectrometry and found no detectable permeation of helium (Fig. 1d). The accuracy was limited only by digital noise of our He spectrometer and a slightly fluctuating background, which yielded $Pr < 10^{-12}$ g/cm$^2 \times$s$\times$bar. Using hydrogen mass spectrometry, no permeation was found either, albeit the accuracy was 3 orders of magnitude lower than for He, because of a larger background. A 12-µm-thick film of PET (polyethylene terephthalate) was used as a reference barrier and exhibited a He leakage rate 1000 times higher than our detection limit (Fig. 1d) yielding PET's bulk permeability $\prod_{He} = Pr \cdot h \approx 10^{-11}$ mm$\times$g/cm$^2 \times$s$\times$bar, in agreement with literature values. The measurements set up an upper limit on $\prod_{He}$ of GO laminates as $\approx 10^{-15}$ mm$\times$g/cm$^2 \times$s$\times$bar, that is, our submicrometre-thick films provide a higher gas barrier than 1-millimeter-thick glass [7].

To evaluate the permeation barrier for liquid substances, we employed weight loss measurements. Fig. 2 shows examples for evaporation rates from a metal container with an aperture covered by a 1-µm-thick GO membrane. No weight loss could be detected with accuracy of <1mg for ethanol, hexane, acetone, decane and propanol in the measurements lasting several days. This sets an upper limit for their $\prod$ as $\approx 10^{-11}$ mm$\times$g/cm$^2 \times$s$\times$bar. Unexpectedly, we observed a huge weight loss if the container was filled with water. Moreover, the evaporation rate was practically the same as in the absence of the GO film, that is, for the open aperture (Fig. 2a). The latter was confirmed directly by using the same aperture with and without a GO cover. Furthermore, the same membrane could be used many times for different liquids, always exhibiting unimpeded and zero evaporation for H$_2$O and other molecules, respectively. Also, after measurements with water, we checked the membranes for a He leak and found none. Only if we increased $h$ to several µm, we could observe a partial inhibition of water evaporation from the container [6], which yielded $\prod_{H2O} \approx 10^{-5}$ mm$\times$g$\times$cm$^{-2}\times$s$^{-1}\times$bar$^{-1}$, that is, water permeates through GO films more than ten orders of magnitude faster than He (Fig. 2b).

To elucidate the origin of the observed 'superpermeability' of water vapor through otherwise leak-tight GO films, we have carried out a number of additional experiments. First, we reduced GO by annealing it at 300°C in a hydrogen-argon atmosphere [4]. The membranes became fragile and required extreme care to avoid cracks but nonetheless became 100 times less permeable to water (Fig. 2a). This can be attributed to structural changes such that $d$ decreased from $\approx 10$ to 4Å, as found by X-ray analysis and in agreement with earlier reports [8,9]. The importance of the interlayer distance was also witnessed when the partial pressure of water was reduced by using salts with calibrated humidity [6]. As $\Delta P$ dropped below 10 mbar, the permeation stopped, which again can be explained by changes in $d$ in low humidity. X-ray analysis shows [10] that this blockage occurs if $d$ decreases below ≈7Å. The process of opening and closing the GO capillaries was found reversible with varying humidity [6].

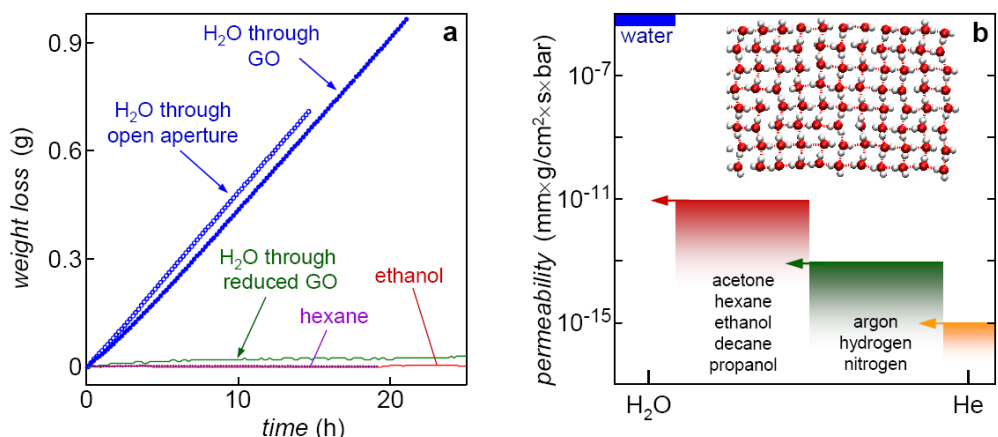

Figure 2. Permeation through GO. **a** - Weight loss for a container sealed with a GO film ($h$ ≈1μm; aperture's area ≈1cm$^2$). No loss was detected for ethanol, hexane, etc. but water evaporated from the container as freely as through an open aperture (blue curves). The measurements were carried out at room temperature ($T$) in zero humidity. **b** - Permeability of GO paper with respect to water and various small molecules (arrows indicate the upper limits set by our experiments). The upper drawing schematically shows the structure of monolayer water inside a graphene capillary with $d$ =7Å, as found in our MDS [6] [also, see refs. 11, 12].

In another series of experiments, we investigated why water permeated through GO film as fast as through an open aperture. To this end, membranes were placed on a support grid that allowed us to apply a water pressure of several bars without damaging them. The large $\Delta P$ did not result in any noticeable increase in $Pr$ with respect to water vapor. On the other hand, if we increased humidity outside the container, $Pr$ decreased. Furthermore, if we blew air at the GO membrane, this notably increased the weight loss rate. Also, $Pr$ increased if the container was heated (we could increase $T$ up to 40°C, above which the freestanding membranes had a tendency to develop cracks). The same changes in $Pr$ happened when we changed $T$ of the membrane only, without heating water inside. In all the cases, $Pr$ changed similarly to the evaporation rate from an open water surface under similar conditions. This suggests that permeation of water through our membranes is limited by evaporation from the wetted surface of GO.

To explain our findings, we propose the following model. Our laminates consist of GO crystallites stacked on top of each other (Figs. 1c). The hydroxyl, epoxy, etc. groups attached to graphene sheets are known [8-10] to be responsible for keeping the relatively large spacing $d$ of about 10Å. It has also been shown that such groups tend to cluster leaving large, percolating regions of a graphene sheet pristine [4,13,14]. Therefore, GO laminates are likely to have an 'empty' space between graphene sheets in the regions free from the adsorbates, which is filled with intercalating water [4,10]. Because $d$ in reduced GO is ≈4Å, the empty space's width $\delta$ can be estimated as ≈5Å, which is sufficient to accommodate a monolayer of water [11,12]. The interlayer water is also known [4,10] to escape from GO in low humidity, leading to smaller $d$ as discussed above and in [6]. We speculate that, if the interlayer water evaporates, the two-dimensional (2D) capillaries adjust their width $\delta$ accordingly and, eventually, become sealed disallowing either water or other molecules to permeate.

The above model allows us to invoke a water permeation mechanism similar to that suggested for small-diameter nanotubes [15-23]. It was argued that they allowed an ordered file of water molecules to move with little friction inside graphene cylinders. To extend this argument into our 2D case, we modeled GO laminates as a network of 2D capillaries (Fig. 1c) and used molecular dynamics simulations (MDS) [6]. Because a graphene monolayer is essentially impermeable, molecular transport in GO laminates should involve a permeation path around graphene platelets. The bottleneck in this process is the passage between graphene sheets separated by $d << L$. For $d \leq 6$Å (including the van der Waals thickness of graphene), our MDS show that water cannot fill the capillaries. On the other hand, for $d \geq 10$Å two layers of water start forming between the sheets. For intermediate $d$, water rushes into the capillaries and forms a highly ordered monolayer shown as the inset in Fig. 2b, in agreement with the previous MDS for 2D hydrophobic capillaries [11,12]. Note that, although graphene is assumed hydrophobic, this concept fails at the monolayer scale [19,20] and our MDS capillaries behave as if they were highly hydrophilic. Furthermore, MDS enabled us to estimate the involved capillary pressures as of the order of 1000 bars [6], in qualitative agreement with the estimates based on van-der-Waals interactions between water and graphite [24]. Such capillary pressure can explain why the water permeation in our experiments was insensitive to $\Delta P$ of several bars. Similar to the case of nanotubes, our MDS monolayer water can move anomalously fast, with velocities reaching m/s and, thus, sufficient to sustain the observed permeation rates [6].

In conclusion, unimpeded evaporation of water through He-leak-tight membranes sounds next to impossible. The closest analogy is probably the permeation of protons (atomic hydrogen) through thin films of transition metals, the phenomenon known as superpermeability [25]. To explain our experiments, we propose the model that can be summarized as follows. GO laminates contain 2D capillaries that, under ambient conditions, are filled with an ordered monolayer of water. A capillary-like pressure provides a sufficient flow to keep the exposed GO surface wet so that the observed permeability is effectively limited by the surface evaporation [26]. Permeation of other molecules is blocked by the intercalating water and, simultaneously, by their shrinkage in low humidity. Such highly selective membranes can be used for filtration and separation. The results have implications for the use of graphene oxide in various applications (e.g., batteries), explaining why the observed surface areas are close to the theoretical maximum. The next challenge is to utilize the found phenomenon, possibly along the lines extensively discussed for membranes made from carbon nanotubes [15-20,27,28].

## Supplementary Material
*Unimpeded permeation of water through helium-leak-tight graphene-based membranes*
R. R. Nair, H. A. Wu, P. N. Jayaram, I. V. Grigorieva, and A. K. Geim

# 1. Fabrication of GO membranes

Graphite oxide was prepared from natural graphite flakes by treating them with potassium permanganate and sodium nitrate in concentrated sulphuric acid[1]. Individual graphene oxide (GO) sheets were exfoliated by dissolving graphite oxide in water with the help of ultrasound, and bulk residues were then removed by centrifugation. To fabricate GO membranes, we used the above suspension to spin- or spray-coat a 25-µm-thick copper foil. To increase the deposition rate, the disks were usually heated to ≈50°C. Freestanding GO membranes of ∼1cm in diameter were prepared by etching away a central part of the copper foil in nitric acid. Finally, the membranes were cleaned in distilled water and dried on a hot plate (<50°C). Steps involved in the fabrication process are schematically shown in Fig. S1. Furthermore, we could produce freestanding membranes by vacuum filtration of the GO suspension through Anodisc filters (0.2 µm pore size). The latter membranes were used in the experiments to apply a high water pressure and for X-ray analysis.

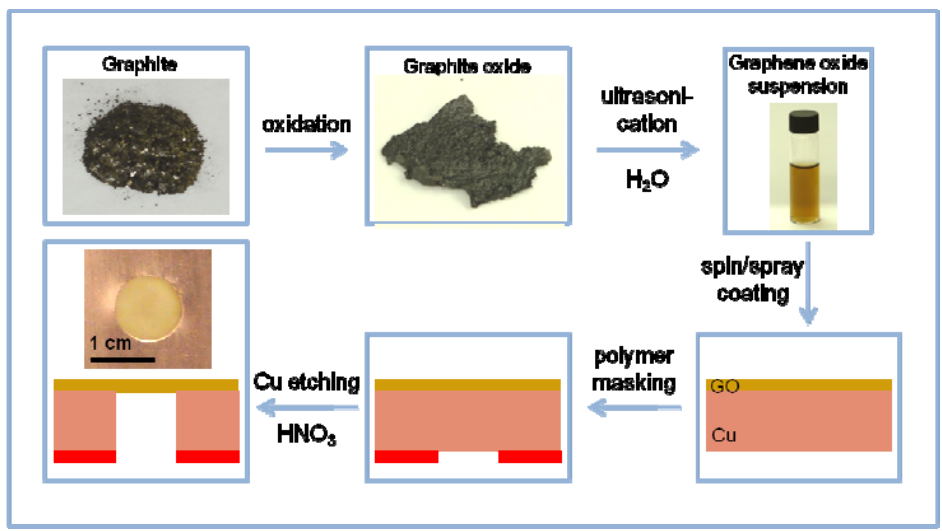

Figure S1. Fabrication procedures for GO membranes.

#2. Experimental setup

Metal containers were fabricated from an aluminium alloy and sealed by using two O rings that clamped the Cu foil from both sides, as shown schematically in Fig. S2. The studied GO membranes covered an aperture in the foils. For gravimetric measurements, we took special care to minimize container's mass and monitored the weight loss by using a computer-controlled ADAM precision balance (accuracy 1mg). To calculate the upper limits for permeability of the studied liquids (shown in Fig. 2b), we used their partial pressures at room *T*.

For mass spectrometry and pressure monitoring, we made larger containers that incorporated inlets with the standard vacuum flanges to allow pumping, pressure gauges and controllable supply of gases (Fig. S2). As a mass spectrometer, we used helium-leak detector INFICON UL200 which allowed detection of helium and hydrogen.

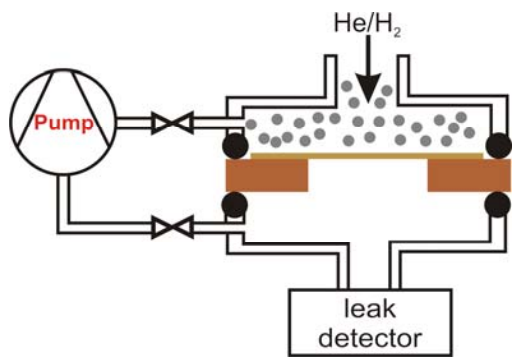

Figure S2. Schematic presentation of the setup used for helium and hydrogen mass spectroscopy. In the permeation measurements by the gravimetric method, we used a light-weight container without extra inlets.

#3. Influence of humidity on water permeation through GO membranes

We have also studied dependence of the evaporation rate on water's partial pressure $\Delta P$. At room $T$, 100% relative humidity (RH) corresponds to ≈23 mbar of water vapour. We could change RH (and, hence, $\Delta P$ for water) inside our sealed container by using various saturated salt solutions[2]. For each salt and for distilled water, we measured the permeation by the weight loss method.

The salts we used were potassium chloride that provides 85% RH, sodium chloride (75%), magnesium nitrate (55%), potassium carbonate (43%), magnesium chloride (33%), potassium acetate (23%) and lithium chloride (11%). To maintain zero humidity outside the container, all the experiments were carried out in a glove box with a negligible water pressure (<$10^{-3}$ mbar). Figure S3 shows the influence of RH (that is, differential vapour pressure $\Delta P$) on water permeation through one of our thin GO membrane ($h$ ≈0.5μm). For comparison, we also plot the measured evaporation rate for the same aperture but in the absence of the GO film (it was mechanically removed).

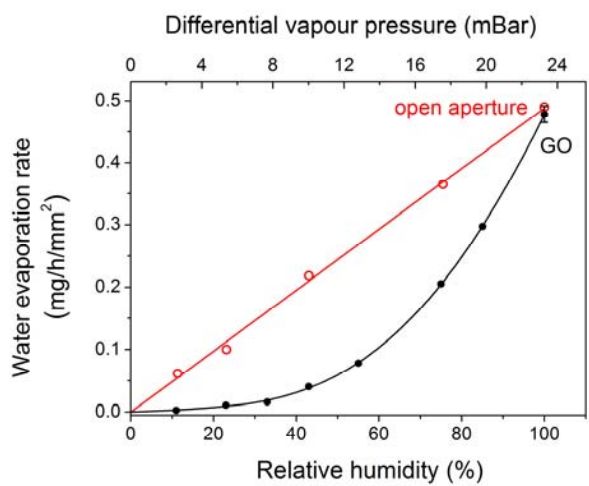

Figure S3. Water permeation through an open aperture and the same aperture covered with a 0.5-μm-thick GO film. Solid lines are guides to the eye. At 100% RH, the permeations rates nearly coincide for this particular thickness but become notably different for thicker membranes (see below).

For 100% RH, both open aperture and GO membrane exhibit nearly the same permeation rate, as shown in Fig. 2 of the main text. However, for lower humidity the permeation rate through GO starts showing strong deviations from that through the open aperture. In the latter case, the evaporation rate decreases linearly with decreasing RH, as expected. In contrast, the GO membrane exhibits a strongly nonlinear behaviour and, for RH below 50%, the water permeation rate through the membrane becomes much smaller than through the same open aperture. The permeation stops at RH below 20% (Fig. S3). The RH required for the stoppage varied from sample to sample and, for some membranes, it stopped at RH as high as 30%. The blocked (dried-out) membranes fully recovered to their initial highly-permeable state after a prolonged (> 1 day) exposure to 100% humidity.

The blockage induced by low RH can be attributed to changes in GO's interlayer spacing $d$, similar to the GO membranes reduced by annealing, in which case $d$ becomes <4Å (see the main text). Indeed, it was previously reported[2] that graphite oxide exhibited a decrease in $d$ with decreasing RH such that $d$ changed from ≈11 to 7 Å for RH varying from 100 to 30%, respectively. To confirm this effect for our films, we have carried out their X-ray analysis at various humidity and found $d$ close to the values reported in ref. 2. It is clear that RH controls the interlayer separation in the GO membranes. For 100% humidity, the observed separation of the order of 10Å should be sufficient to accommodate one monolayer of water molecules and allow their diffusion. For lower RH, this freedom of movement becomes restricted because of the smaller $d$. As more and more capillaries dry out and become effectively closed, the water permeation eventually stops (see the main text).

#### #4. Length of 2D capillaries and flow velocity
Figure 1c of the main text shows schematically the suggested permeation path for water through GO membranes. They have a layered structure made of graphene sheets separated by an interlayer distance $d$ between ≈7 and 11Å, depending on the humidity (see #3). Taking into account that the electronic clouds around graphene sheets extend over a distance of $a$ ≈3.5Å, the above separation $d$ translates into an 'empty' space of width $\delta = d - a$ =3 to 7Å, which can be available for other molecules to diffuse (see Fig. 1c of the main text). With reference to this figure, the complete path of water through a GO film of thickness $h$ involves a number of turns $N = h/d$, and each turn involves a capillary length $L$ ~1µm (we assume that flakes have the same typical width and length). Therefore, the total length $l$ of each effective channel is given by $N \times L$, yielding $l$ ~1mm for a typical 1-µm-thick membrane.

The total number of the channels per unit area can be estimated as by $1/L^2$, that is, our membranes with a typical area of ≈1cm$^2$ effectively contain ~10$^8$ parallel 2D channels. Nanocapillaries occupy a relatively small part of the total membrane area ($\pi D^2/4$) and this fraction $\nu$ can be estimated as $\delta/L$ ~0.1%. To sustain the observed evaporation rate $Q$ ≈50mg/h per cm$^2$ at room $T$ (Fig. S4), it requires a water flow inside the capillaries with velocity $V = Q/\nu\rho$ where $\rho$ is the water density. This yields $V$ of the order of 0.01 cm/s.

#### #5. Effect of membrane's thickness on water permeation
The observation of the same evaporation rates from an open container and the same container covered with a 0.5-µm-thick GO membrane (Figs. 2a, S3) poses an important question: at what thickness does the presence of the GO films start playing any role? To answer this question, we employed GO membranes with $h$ up to 10 µm, the maximum thickness that we could achieve by spray/spin-coating. Figure S4 shows water loss rate $Q$ as a function of $h$. The measurements were carried out by the same gravimetric method under the identical conditions. It is clear that the vapour barrier is practically absent for the submicron GO membranes but gradually builds up for thicker films. For our thickest membrane, the flow is impeded by a factor of two. The observed changes in the evaporation rate have allowed us to estimate the permeation for bulk GO as stated in the main text.

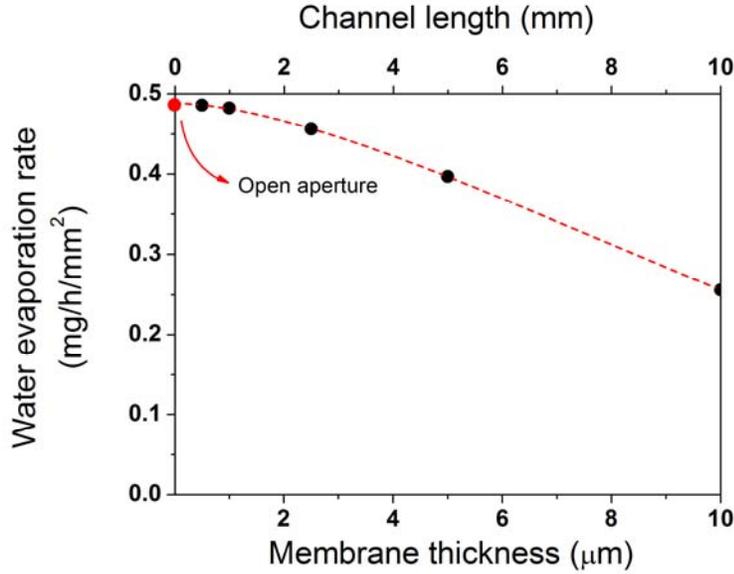

Figure S4. Dependence of the evaporation rate on GO membrane's thickness (symbols). The dashed curve is a guide to the eye. The upper axis shows the effective channel length *l* that was calculated as described in section #4.

#### #6. Comparison with classical flow equations

It is instructive to compare the experimentally observed permeation rate with that expected from the classical equations for the given sizes of nanocapillaries. If we assume that water vapour permeates through our GO membranes as a gas and, therefore, the Knudsen regime is applicable (diffusion is limited by collisions with graphene walls), the mass flow per unit area can be estimated as

$$Q \approx \delta^3 \left(\frac{1}{L^2}\right)\left(\frac{M}{RT}\right)^{1/2}\left(\frac{\Delta P}{l}\right)$$

where $M$ is the molecular weight of water, $R$ the gas constant and $1/L^2$ the area density of 2D channels. For simplicity, let us consider our thickest membrane ($h \approx 10$ μm and the effective channel length $l \sim 10$ mm), in which the observed weight loss becomes limited by GO properties rather than the aperture's area (Fig. S4). At this thickness, the membranes allow flux $Q \sim 25$ mg/h per cm$^2$ (see #5). On the other hand, the Knudsen formula for such membranes yields $Q \sim 10^{-9}$ mg/h per cm$^2$. This is ten orders of magnitude smaller than the water vapour flux observed experimentally.

Alternatively, assuming that water inside our nanocapillaries behaves as a classical liquid, we can employ the Hagen-Poiseuille equation

$$Q \approx \delta^3 \left(\frac{1}{12\eta}\right)\left(\frac{1}{L}\right)\left(\frac{\Delta P}{l}\right)\rho$$

where $\eta$ is the viscosity of bulk water (1 mPa·s) and $\rho$ its density. For the other parameters remaining the same, the latter equation yields $Q \sim 10^{-6}$ mg/h per cm$^2$, that is, nanoconfined water exhibits a flow enhanced by a factor $\Theta \sim 10^7$ with respect to the classical laminar regime. In both equations above, we have assumed the differential pressure $\Delta P$ to be 23 mbar (100% RH inside the sealed containers). However, monolayer water interacts with the graphene surface and, according to simple van-der-Waals-energy considerations,[3] this interaction should lead to a capillary-like pressure $P$ of $\approx 10^3$–$10^4$ bar. This is in conceptual agreement with our molecular dynamics simulations (MDS) below, which yields $P$ of the order of $10^3$ bar. Therefore, it is appropriate to use this value as the driving pressure $\Delta P$ in the Hagen-Poiseuille equation. Then, our experiments infer an enhancement factor $\Theta$ of only a few hundred, in good agreement with the enhancement factor that was recently[4,5] reported for the case of sub-1-nm nanotubes with one file of moving water. Furthermore, the enhanced water flow in nanoporous materials is often described by the slip length $l_s$, a parameter describing the lowered friction between water and capillary walls. Our experiments allow an estimate for $l_s \sim \Theta \cdot \delta/8 \approx$ 10-100 nm, again in agreement with the most recent measurements for the correlated water flow in the 1D geometry[5].

#### #7. Atomistic simulations of water dynamics in 2D graphene capillaries

To gain a theory insight about water permeation through 1-nm 2D capillaries, we employed classical molecular dynamics simulations (MDS), which have previously been proven an efficient tool to investigate molecular transport at nanoscale. Our MDS were carried out by using the LAMMPS software package from Sandia National Laboratories[6]. Figure S5 shows the setup that consists of two graphene reservoirs and a connecting 2D graphene capillary. All the carbon atoms in the model are assumed static. For water simulations, we have employed the SPC/E model[7] with the O-H bond length of 1.0Å and the H-O-H angle of 109.47°. The harmonic style was used for both bond and angle potentials (see LAMMPS Manual). The charges on the oxygen site and the hydrogen sites were chosen –0.8476 $e$ and +0.4238 $e$, respectively ($e$ is the free electron charge). The electrostatic interaction was modelled by using the Coulomb potential. Van der Waals interactions between atoms were described by a 12-6 Lennard-Jones (LJ) potential with parameters $\varepsilon_{O-O}$ =0.1553 kcal/mol and $\sigma_{O-O}$ =3.166Å. The particle–particle particle–mesh solver was applied to account for the truncation of the long-range electrostatic forces.[6] The LJ potential parameters for the C-O interaction were calculated by the Lorentz-Berthelot combining rules, and we used the same parameters as in ref. 8, that is, $\varepsilon_{C-C}$ =0.0553kcal/mol, $\sigma_{C-C}$ =3.4Å, $\varepsilon_{C-O}$ =0.0927kcal/mol and $\sigma_{C-O}$ =3.283Å. The cut-off distance for all LJ potentials was chosen 10Å. For computational efficiency, LJ parameters for the atomistic interactions involving hydrogen atoms were set to zero. The simulations were carried out for $T$ =300K.

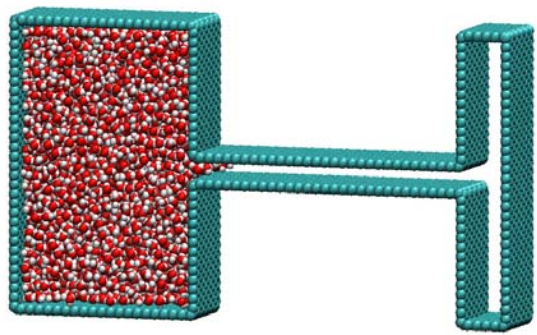

Figure S5. Model used to simulate water-molecule dynamics in 2D graphene nanocapillaries.

To account for the dimensionality ($L \gg d$), we used the periodic boundary conditions with a period of $\approx$21Å (out-of-plane direction in Fig. S5). The length of the 2D capillary in Figure S5 is $\approx$74Å (30 graphene hexagons). The length of the left reservoir is $\approx$50Å and its height $\approx$75Å. MDS were carried out for 4 different widths $d$ of the 2D slit: namely, 6, 6.5, 7 and 7.5Å (measured between the centres of carbon atoms). A fixed number of water molecules were initially put in the left reservoir, while the capillary and the right reservoir were kept empty. If a water molecule passed through the channel, it disappeared ('evaporated') at the other end, in the right reservoir.

The number of water molecules entering the right reservoir is shown as a function of time in Figure S6. Our MDS revealed that water molecules tended to enter the capillary and initially moved through it with a velocity of $\approx$20m/s (Fig. S6). They reached the evaporation point after ~0.4 ns for all the channels with $d$

>6Å. On the other hand, the 6Å channel did not allow any noticeable permeation. This indicates that $d$ >6Å is necessary to accommodate a single layer of water molecules between MDS' graphene sheets.

The initial density of water in the reservoir was chosen larger than that of bulk water. Still, the differential pressure could not cause any flow through the 6Å slit. For the wider capillaries, the density of water molecules in the left reservoir gradually decreased because of the evaporation process and, therefore, the differential pressure decreased too. This is why water molecules moved fast during the first 2 ns and then slowed down (Fig. S6). Importantly, after 5ns, the water density in the left reservoir became close to or smaller than that of bulk water. In this regime, the initial hydrostatic pressure is no longer the driving force but water molecules still fill in and move through the capillary.

The fact that the water fills the 2D channel even under a negative pressure in the left reservoir indicates an additional driving mechanism that can be attributed to the interaction between monolayer water and graphene walls in the capillary, as discussed in #6. We have carried out further MDS to estimate this effective capillary pressure $P$. To this end, simulations were restarted from molecular configurations reached after 10ns, when the channel was fully filled, and an additional force was applied to all oxygen atoms in the direction against the flow. This mimics a gravitational force and can directly be translated into an extra capillary pressure $P$ in the slit. We found that if $P$ exceeded ≈500 bars, water molecules in the channel were pulled back to the left reservoir, and the 2D capillary started to dry out. The found value of $P$ in our 1-nm-sized slit is in qualitative agreement with the classical estimates[3] using the van der Waals interaction between water and graphite.

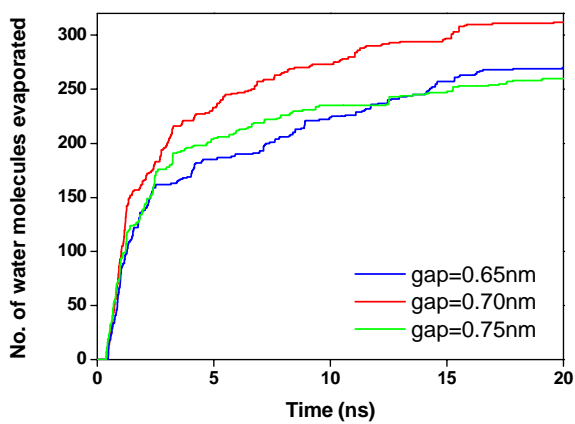

Figure S6. The number of water molecules reaching the right reservoir as a function of time. Different colours correspond to different channel widths.